\documentclass{emulateapj}

\usepackage{color}
\usepackage{rotating}
\shorttitle{Dark matter subhalos in UMi}
\shortauthors{Lora et al.}

\begin{document}


\title{Dark matter subhalos \\
    in the Ursa Minor dwarf galaxy}


\author{V. Lora\altaffilmark{1}, A. Just\altaffilmark{1}, 
F. J. S\'anchez-Salcedo\altaffilmark{2} and E. K. Grebel\altaffilmark{1}}
\altaffiltext{1}{Astronomisches Rechen-Institut, Zentrum f\"{u}r Astronomie der Universit\"{a}t Heidelberg, \\
             M\"{o}nchhofstr. 12-14, 69120 Heidelberg, Germany}
\altaffiltext{2}{Instituto de Astronom\'{\i}a,
              Universidad Nacional Aut\'onoma de M\'{e}xico,
              AP 70-264, 04510 D.F., M\'{e}xico}
\email{vlora@ari.uni-heidelberg.de}


% \altaffiltext{1}{Visiting Astronomer, Cerro Tololo Inter-American Observatory.
% CTIO is operated by AURA, Inc.\ under contract to the National Science
% Foundation.}
% \altaffiltext{2}{Society of Fellows, Harvard University.}
% \altaffiltext{3}{present address: Center for Astrophysics,
%     60 Garden Street, Cambridge, MA 02138}
% \altaffiltext{4}{Visiting Programmer, Space Telescope Science Institute}
% \altaffiltext{5}{Patron, Alonso's Bar and Grill}


\begin{abstract}
Through numerical simulations, we study the dissolution timescale
of the Ursa Minor cold stellar clump, due to the combination of 
phase-mixing and gravitational encounters with compact dark substructures
in the halo of Ursa Minor. 
We compare two scenarios; one where the dark halo is made up by a smooth
mass distribution of light particles and one where the halo contains $10\%$ 
of its mass in the form of substructures (subhalos). 
In a smooth halo, the stellar 
clump survives for a Hubble time provided that the dark matter halo has
a big core.
In contrast, when the point-mass dark substructures are added, 
the clump survives barely for $\sim 1.5$~Gyr. These results 
suggest a strong test to the $\Lambda$-cold dark matter scenario at
dwarf galaxy scale.
\end{abstract}



\section{Introduction}

The $\Lambda$ Cold Dark Matter ($\Lambda$CDM) model has been proved to be 
successful in reproducing structure formation on large scales.
%\textbf{
In the standard paradigm, a nearly scale invariant power spectrum
describes the cosmological primordial density fluctuations.
If DM is collisionless, the clustering
of substructure is scale-invariant down to the free-streaming 
scale of the CDM particle (e.g., \citeauthor{hofmann01}~\citeyear{hofmann01}; 
\citeauthor{green05}~\citeyear{green05}).
The properties of the subhalos in Milky Way size galaxies have been studied 
by \citet{gao04}, \citet{springel08} and
\citet{diemand08} through collisionless DM simulations.
%}
They found that $10$\% of their mass reside  in 
$\sim3\times10^{4}$ subhalos with masses in the range of $10^{4} - 
10^{8} M_{\odot}$, following a CDM subhalo-mass function 
$dN/dM \propto M^{-1.9}$. 

%\textbf{
At scales of dwarf satellite galaxies, a comparison
between the model predictions and observations is limited by 
our poor understanding of the baryonic
processes involved in the formation of galaxies.
For instance, supernova feedback and gas heating from 
cosmic sources can alleviate
the inconsistency between the observed number
of satellite galaxies in the Milky Way halo and the much 
higher number of subhalos predicted using 
$N$-body $\Lambda$CDM simulations 
\citep{moore99,klypin,ben02,som02,ostriker,fon11}.
In this model,
many satellites are either too faint to be detected in surveys
or dark matter subhalos devoid of a baryonic counterpart \citep{boylan,bovill}.
%}

Some observational consequences of the existence of substructure 
within galactic halos have been studied in the literature.
\cite{romano08} suggested that long-lived stellar bars can be triggered by
a tide from a massive subhalo.
\cite{kannan12} explored the idea that the interaction of dark
matter subhalos with the gaseous disks of galaxies may generate density
enhancements in the gas.
\citet{johnston02} studied the distribution of carbon stars in the
stream of the Sagittarius (Sgr) dwarf galaxy and found that the
Sgr debris is more strongly scattered than would be expected if it would be 
orbiting in a smooth (and thus non-substructured) DM halo, 
but that is entirely 
consistent with perturbations by the Large Magellanic Cloud (LMC) alone.
In this line of research, \citet{carlberg09} modeled the interaction 
between streams and
halo clumps and concluded that stellar streams older than $3$~Gyr 
cannot survive in the presence of subhalos with the masses and numbers 
predicted by the $\Lambda$CDM model. 
Nevertheless, the density variations in the star stream detected around the
globular cluster Palomar 5 and in the galaxy M31  
appear to be in agreement with the existence of DM subhalos
\citep{yoon11,carlberg11}.

The presence of dark-matter subhalos in dwarf spheroidal galaxies (dSph)
might have observable consequences and may shed light on
the nature of DM particles.  For instance,
\citet{jin05} and \citet{totani10} suggested that compact massive 
dark objects in the halo of dSph galaxies 
could resolve some long-standing problems in these systems. 
On the other hand,
\cite{penarrubia10} used analytical and $N$-body methods to examine the survival of
wide stellar binaries against encounters with dark subhalos
orbiting in the DM halos of dwarf galaxies. They found that a large fraction of
wide binaries can be wiped out due to tidal encounters with these dark substructures. 
The observations of large separation binaries would impose a strong test to the putative 
substructure in halos of dwarf galaxies. 

%************************  EDITOR
Also \cite{ma04} studied the gravitational scattering effects caused by subhalos 
on the phase-space distribution of DM particles in main halos. Their numerical 
experiments indicate that the number and the mass density of subhalos could be high 
enough to flatten the main halo's inner cusp within a few dynamical times.
%************************

In this work, we will focus on dSph galaxy Ursa Minor (UMi).
UMi has the peculiarity of showing
a second stellar density peak (or clump) which is believed to be a cold 
long-lived structure \citep{palma03}. 
This clump has survived because the underlying 
gravitational potential in UMi must be close to harmonic, which
can be accomplished if the density profile of the dark halo has a 
large core  \citep{kleyna03}. 
In addition, \cite{sanchez07} established stringent constraints on the 
mass and abundance of compact objects in UMi 
in order to preserve the
integrity of cold small-scale clumps seen in some dSph galaxies.
In the present work, our aim is to test whether or not the putative
subhalos with a mass spectrum as that found in collisionless CDM
is consistent with the stellar clump found in UMi.
Our work may shed light on 
the mechanism responsible for the formation of cores in dwarf galaxies
because some of them predict dissolution of dark-matter substructures
below a scale of $\sim 1$ kpc. In fact,
the core-making mechanisms proposed so far fall in three
broad categories: supernova feedback (e.g., Mo \& Mao 2004;
Mashchenko et al.~2006; Governato et al.~2010, 2012; Pontzen \&
Governato 2012; Macci\`o et al.~2012), dynamical friction from infalling
baryonic clumps (El-Zant et al.~2001; 
Romano-D\'{\i}az et al.~2009; Goerdt et al.~2010), 
or a change in any of the basic properties of dark matter
particles [e.g., collisional, annihilating, decaying or warm dark matter]
(Spergel \& Steinhardt 2000; Kaplinghat, Knox \& Turner 2000; Cen 2000;
S\'anchez-Salcedo 2003; \'Avila-Reese et al.~2001).
Models in the latter category are expected to help erase dark-matter 
subhalos in dwarf galaxies,
e.g., by mass stripping if dark matter is collisional or by suppression of
the power in the mass spectrum at small scales if dark matter is warm,
whereas models in the first two categories are expected to preserve them.

This article is organized as follows.
In \S \ref{sec:UMI} we describe some properties of UMi and its clump.
The initial conditions for our $N$-body simulations are described
in \S\ref{sec:dark}. The results of the simulations are shown in
\S \ref{sec:results}. Finally, we present our conclusions
in \S\ref{sec:conclusions}.

%%%%%%%%%%%%%%%%%%%%%%%%%%%%%%%%%%%%%%%%%%%%%%%%%%%%%%%%%%%%%%%%%%%%%%%%%%%%%%%

\section{UMi and its clump}
\label{sec:UMI}
UMi is a dSph galaxy satellite of the Milky Way, located at a Galactocentric 
distance of $R_{GC}=69 \pm 4$~kpc \citep{grebel03}.
Dynamical studies suggest that UMi has a
mass-to-light ratio larger than $60 M_{\odot}/L_{\odot}$ 
(e.g.~\citeauthor{wilkinson04}~\citeyear{wilkinson04}).
Rescaling the results of \citet{gilmore07}
and taking a luminosity of $L_{V}=1.1\times10^{6}L_{\odot}^{V}$
\citep{palma03}, the total mass within $0.6$~kpc is 
$6.3\times10^7 M_{\odot}$, which results in a mass-to-light ratio 
 $M/L_{V} \gtrsim 60 M_{\odot}/L_{\odot}$ for UMi. It has to be noted that if we take 
$L_{V}=3\times10^{5} L_{\odot}$ \citep{grebel03} then we obtain a  
$M/L_{V} \sim 200 M_{\odot}/L_{\odot}$ \citep{gilmore07}. Deeper
data for dwarf galaxies often reveal a larger angular extent of their 
stellar component (e. g., \citeauthor{odenkirchen} \citeyear{odenkirchen};
\citeauthor{kniazev} \citeyear{kniazev}); nevertheless, 
we adopt the lower luminosity value for UMi.
The high value of $M/L$ in UMi implies that it is one of the most 
dark matter dominated dSph galaxies in the Local Group. 

UMi's stellar King core radius along the semimajor axis is 17.9 arcmin 
($\sim0.4$~kpc) \citep{palma03}. UMi shows 
a large ellipticity in the shape of the inner isodensity contours of the 
surface density of stars ($\epsilon =0.54$) \citep{palma03}, but the most 
remarkable feature in UMi's structure is the second off-centered density 
peak \citep{kleyna98}. One of these two peaks is located on the north-eastern side 
of the major axis of UMi at a distance of $\sim 0.4$ kpc from UMi's center.
The radial velocity distribution of the stars in UMi is well fitted
by two Gaussians, one representing the underlying background ($8.8$~km s$^{-1}$) and 
the other representing the velocity dispersion of the second peak ($0.5$~km~s$^{-1}$). 
The stars in the vicinity of this peak comprise a kinematically distinct 
cold subpopulation: a dynamically cold stellar clump.
The most appealing interpretation of UMi's clump is that it is a disrupted 
cluster \citep{read06} that orbits in the plane of the sky,
and that has survived in phase-space because the 
underlying gravitational potential is harmonic \citep{kleyna03}. This implies
that the dark halo density profile in UMi should have a core 
(and not a cuspy profile
as predicted by the $\Lambda$CDM paradigm) and that this core must
be large ($\sim500$~pc, see \citeauthor{kleyna03}~\citeyear{kleyna03} and 
\citeauthor{lora}~\citeyear{lora}).

% Sobre el gas
%\textbf{
The stellar population of the UMi dwarf galaxy is very old with an age of
$10$--$13$ Gyr; virtually all the stars were formed $10$~Gyr ago, and 90\%
of them were formed $13$~Gyr ago \citep{carrera02,rocha11}. 
This implies that star formation was halted at a
redshift of $z\sim 2$, probably because supernova explosions 
were able to remove all the gas.
In fact, deep observations have not detected gas in UMi \citep{young99,young00}.
Gas replenishment from dying stars and stellar winds is expected to be
very small. If the gas returned to the interstellar medium
by intermediate- and low-mass dying stars were
distributed across the galaxy as the stars are, the central
gas density would reach a value of 
$\sim4.4\times10^{-3}M_{\odot}$~pc$^{-3}$ after $10$~Gyr \citep{carrera02}. This
corresponds to a particle density of $5\times10^{-3}$~cm$^{-3}$,
much smaller that the typical lower thresholds for star-forming 
regions ($\sim1000$~cm$^{-3}$) \citep{shu}. 
Since we are interested here in the dynamical evolution of the stellar clump
over the last $10$ Gyr,
once essentially all the gas has been removed,
we can discard both gas processes and stellar
evolution in our simulations.
%}

%%%%%%%%%%%%%%%%%%%%%%%%%%%%%%%%%%%%%%%%%%%%%%%%%%%%%%%%%%%%%%%%%%%%%%%%%%%%%%%%%%%%%%%%%%%%%%

\section{$N$-body: initial conditions and the code}
\label{sec:dark}

\subsection{The dark matter component of UMi}
We performed $N$-body simulations of the UMi dSph galaxy, which include a baryonic component
(the bulk stellar component and the stellar clump) embedded in a
spherical (live) dark matter halo.
The dark matter halo density profile selected is defined as
\begin{equation}
\rho(r)=\frac{\rho_{0}}{(r/r_s)^{\gamma} [1+(r/r_s)^{\alpha}]^{(\beta-\gamma/\alpha)} }  \mbox{   .}
\end{equation}
Here  $\rho_{0}$ accounts 
for the central density in the case of $\gamma=0$ and $r_s$ is the scale length. 
A NFW profile \citep{nfw}
is obtained for ($\alpha,\beta,\gamma$)=($1,3,1$). 
We have chosen the values
($\alpha,\beta,\gamma$)=($1,3,0$) in order to have a cored DM 
mass density profile. 

It has been found that cored profiles 
(and not cuspy profiles) are in better agreement with UMi dynamics
\citep{kleyna03,lora}. Some other dSph galaxies in the 
Local Group such as Fornax, Sculptor, Carina, 
Leo I and Leo II are also believed to possess cored dark halos 
instead of cuspy profiles
%\textbf{
\citep{kleyna03,goerdtd06,goerdtd10,sanchez1,gilmore07,battaglia,
amorisco,walker11,jardel12,agn12}.
%} 
Moreover, it is found from high resolution observations of the rotation curves 
for DM dominated low surface brightness (LSB) galaxies that 
DM halos have density profiles compatible with flat central cores 
\citep{deblok01,chen10}.

We explored two different cored halos;
a small-core halo with a scale length of $0.91$~kpc
(which corresponds to a core radius of $\sim 0.4$~kpc) with a total 
mass of $M=2\times10^9 M_{\odot}$, and a big-core halo with a scale 
length of $2.2$~kpc (core radius $\sim1$~kpc) and a
total mass of $M=3\times10^{10} M_{\odot}$. 
For our halo models we have 
a mass within a radius of $0.39$~kpc (the clump's orbit) 
of $\sim10^7 M_{\odot}$ for the small-core halo, and a mass of
$\sim3\times10^7 M_{\odot}$ for the big-core halo model.
This is consistent with the results of \citeauthor{strigari07} (2007, 2008)
who considered dark halos
compatible with the observed stellar kinematics of the
classical dSph galaxies, including the dwarf galaxy UMi. 
They found that, for realistic
density profiles, the mass interior
to $300$ pc is $\sim 10^{7}M_{\odot}$ for all dSph galaxies in
the Milky Way halo. 

To generate the initial conditions of the dark matter particles of 
the smooth background distribution, we used 
the distribution function proposed by \citet{widrow00}, and 
the velocity dispersion of the system was taken to be isotropic.

In order to explore the dynamical effects of a clumpy halo,
we generated the mass spectrum of substructures in the halo 
following the power-law mass distribution
\begin{equation}
\frac{dN}{dM} = a_{0} \left(\frac{M}{M_{0}}\right)^{-1.9} 
\label{eq:substructure}
\end{equation}
\citep{gao04,springel08,diemand08}.
The values of $a_{0}$ and $M_{0}$
are fixed once the lower $M_{min}$ and upper $M_{max}$ mass limits 
of the substructure particles 
and the fraction ($f$) of the halo mass 
comprised in substructures, are given. 

The mass of dark matter particles used in our simulations 
without subhalos is $2\times10^{3}M_{\odot}$, which corresponds to
our maximal mass resolution for the halo particles. 
Then it is a natural choice to 
take $M_{min}=10^{3.3}M_{\odot}$ as the lower mass limit.
Adopting the subhalo mass function found in CDM simulations, the subhalos 
are expected to have masses up to $M_{max}\lesssim0.1 M_{vir}$,
where $M_{vir}$ is the virial mass of the halo 
\citep{penarrubia10}. 
Given that the estimated average
virial mass of the dSph galaxies is of the order of $M_{vir}\sim10^9M_{\odot}$ 
\citep{penarrubia08,walker09}, the upper mass limit could be as 
large as $10^8M_{\odot}$. 
It then becomes reasonable to take the generous upper mass limit as 
$M_{max}\simeq 10^7M_{\odot}$.

Equation (\ref{eq:substructure}) with $f$, $M_{min}$ and $M_{max}$ known,
determines 
the number $N_{SUB}$ of substructure particles.  
In order to construct a clumpy DM halo, we choose
from the original smooth DM halo, a number $N_{SUB}$ of
random particles and replace their masses according to the power-law
in Equation (\ref{eq:substructure}). 

%%%%%%%%%%%%%%%%%%%%%%%%%%%%%%%%%%%%%%%%%%%%%%%%%%%%%%%%%%%%%%%%%%%%%%%%%%%%%%%%%%%%

\subsection{The baryonic component of UMi}
\label{sec:barion}

The spatial stellar density profiles of elliptical systems are commonly described 
as a power law in radius 
\begin{equation}
\rho_{*}(r)=\frac{(3-\gamma) M_{*}}{4\pi} \frac{a}{r^{\gamma} (r+a)^{\beta-\gamma}},
\end{equation}
where $M_{*}$ is the total stellar mass and $a$ is the scale radius.
For $\beta = 4$, these models have simple analytic properties and are called 
Dehnen models (\citeauthor{dehnen93} \citeyear{dehnen93}; 
\citeauthor{tremaine94} \citeyear{tremaine94}).
In these models the densities are proportional to $r^{-4}$ at large radii and diverge
in the center as $r^{-\gamma}$. 
To model the bulk stellar component in UMi we used a Dehnen model \citep{dehnen93}
where $\gamma=3/2$. This slope most closely resembles the de Vaucouleurs model in surface density.
We took the scale radius for the underlying stellar component in the UMi galaxy to be $a=0.4$~kpc 
\citep{kleyna98,palma03}. We set the total mass of the stellar component to be
$M_{\star}=9\times10^5M_{\odot}$ taking the typical value of the mass-to-light 
ratio $\Upsilon_{\star}=3$. 

We performed an $N$-body simulation with the DM halo and
the underlying stellar components together. The resulting system was 
found to be stationary for a Hubble time (i.e. the density profile of the 
stellar component and the velocity dispersion  
stayed approximately constant for a Hubble time). 
The stellar velocity dispersion 
($\sigma \sim 10$~km~s$^{-1}$) was in agreement with the velocity dispersion of the
stellar component found in UMi ($\sigma=9.3, 12, 9.5\pm1.2$~km s$^{-1}$; 
\citeauthor{wilkinson04} \citeyear{wilkinson04}; \citeauthor{gilmore07} \citeyear{gilmore07}; 
\citeauthor{walker09} \citeyear{walker09}).

Finally, for the initial density profile of the stellar clump, we take 
\begin{equation}
 \rho_{c}(r) = \rho_{0} \exp(-r^{2}/2r_{c}^{2}) \mbox{,}
\end{equation}
with the clump radius $r_{c}$ between $12$~pc and $35$~pc (\citeauthor{palma03}
\citeyear{palma03}). 
The clump's velocity dispersion was set to $1$~km s$^{-1}$. 
The clump was dropped at a galactocentric
distance of $0.39$~kpc in a circular orbit in the $(x,y)$ plane (see 
also \citeauthor{sanchez10} \citeyear{sanchez10}).
Since the clump's stars have the same 
color as the underlying stellar population in UMi \citep{kleyna98}, we  
assume that the $V$-band mass-to-light ratio $M/L_{V}$ of the clump
is the same as it is for the underlying stellar component (say, $\Upsilon_{\star}=3$).
Thus, the mass of the clump $M_{c}$ is $\simeq 4\times10^{4} M_{\odot}$.

%%%%%%%%%%%%%%%%%%%%%%%%%%%%%%%%%%%%%%%%%%%%%%%%%%%%%%%%%%%%%%%%%%%%%%%%%%%%%

\subsection{The code}
\label{sec:code}

%\textbf{
Since the internal two-body relaxation 
timescales for the three components (clump, underlying stellar component 
and halo) are much larger than one Hubble time, 
this system can be represented as collisionless \citep{binney}. 
We simulated the evolution of the UMi dwarf galaxy 
(stellar clump, underlying stellar 
component and DM halo) using the $N$-body code \scriptsize {SUPERBOX} \normalsize \citep{fellhauer00}. 
\scriptsize {SUPERBOX} \normalsize is a highly efficient particle-mesh, 
collisionless-dynamics code with high resolution sub-grids.
The parameters of each model are given in Table \ref{tab:resultados1}.
%}

In our case, \scriptsize {SUPERBOX} \normalsize uses three nested 
grids centered in the center of density of the UMi dSph galaxy. We used $128^3$ cubic
cells for each of the grids. The inner grid is meant to resolve the 
inner region of UMi 
and the outer grid (with a radii of $1000$~kpc for all cases) resolves the stars that are stripped away from 
UMi's potential. 
%\textbf{
The spatial resolution is determined by the number of grid cells per
dimension ($N_c$) and the grid radius ($r_{\rm grid}$). 
Then the side length of one grid cell is defined as
%\begin{equation}
$l=\frac{2 r_{\rm grid}}{N_c-4}$.
%\end{equation}
To study the convergence of the results with different resolutions,
we performed three simulations: a simulation with $128^3$ cubic cells, one with higher ($256^3$ cubic
cells) and one with lower ($64^3$ cubic cells) resolution of model $M3$ 
(see model parameters in 
Table \ref{tab:resultados1}). 
We obtained practically the same results for the
three resolutions. We conclude that $128^3$ cubic cells
suffices to achieve a robust accuracy. 
For $N_{c}=128$, the resolution, which is of  
the order of the typical distance between the particles in the simulation,
is given at the two last columns of Table \ref{tab:code}.
%}
%(for further see \citep{khoperskov}).}

\scriptsize {SUPERBOX} \normalsize integrates the equations of motion with a 
leap-frog algorithm, and a constant time step $dt$. We selected a time 
step of $dt=0.1$~Myr in our simulations in order to guarantee that the
energy (for the isolated components) is conserved better than $1\%$.
The properties of the three modeled components of UMi (DM halo, underlying
stellar component and stellar clump) 
used in the simulation are shown in Table \ref{tab:code}. Also, 
the radii of the inner, middle and
outer grids for each of the components are given.

%%%%%%%%%%%%%%%%%%%%%%%%%%%%%%%%%%%%%%%%%%%%%%%%%%%%%%%%%%%%%%%%

\section{Results}
\label{sec:results}

Our $N$-body simulations were carried out from a time $t=0$ to $t=10$~Gyr using 
the code described in Section \ref{sec:code}. 
In Table \ref{tab:resultados1} we list 
the parameters used in the simulations (see also \S \ref{sec:barion}).
For the models $M1-M4$ (small-core halo) we used $10^6$ particles to represent the halo component, $10^5$ particles
for the bulk stellar component, and $10^4$ particles for the stellar clump. For the models $M5-M8$ (big core halo)
we used the same particle number mentioned before for the underlying
stellar component and for the
stellar clump, but we used $1.5\times10^7$ particles 
for the halo component. This was done in order to have the same mass per particle
in both small ($2\times10^9 M_{\odot}$) and big ($3\times10^{10} M_{\odot}$) 
core DM halos.
In total we have a set of $8$ simulations (see $M1$-$M8$ in Table~\ref{tab:resultados1}).

%%%%%%%%%%%%%%%%%%%%%%%%%%%%%%%%%%%%%%%%%%%%%%%%%%%%%%%%%%%%%%%%%%%%%%%%%%%%%%%%

\subsection{Halo without substructure}
\label{sec:sin_sub}

\begin{figure}
\epsscale{.80}
\plotone{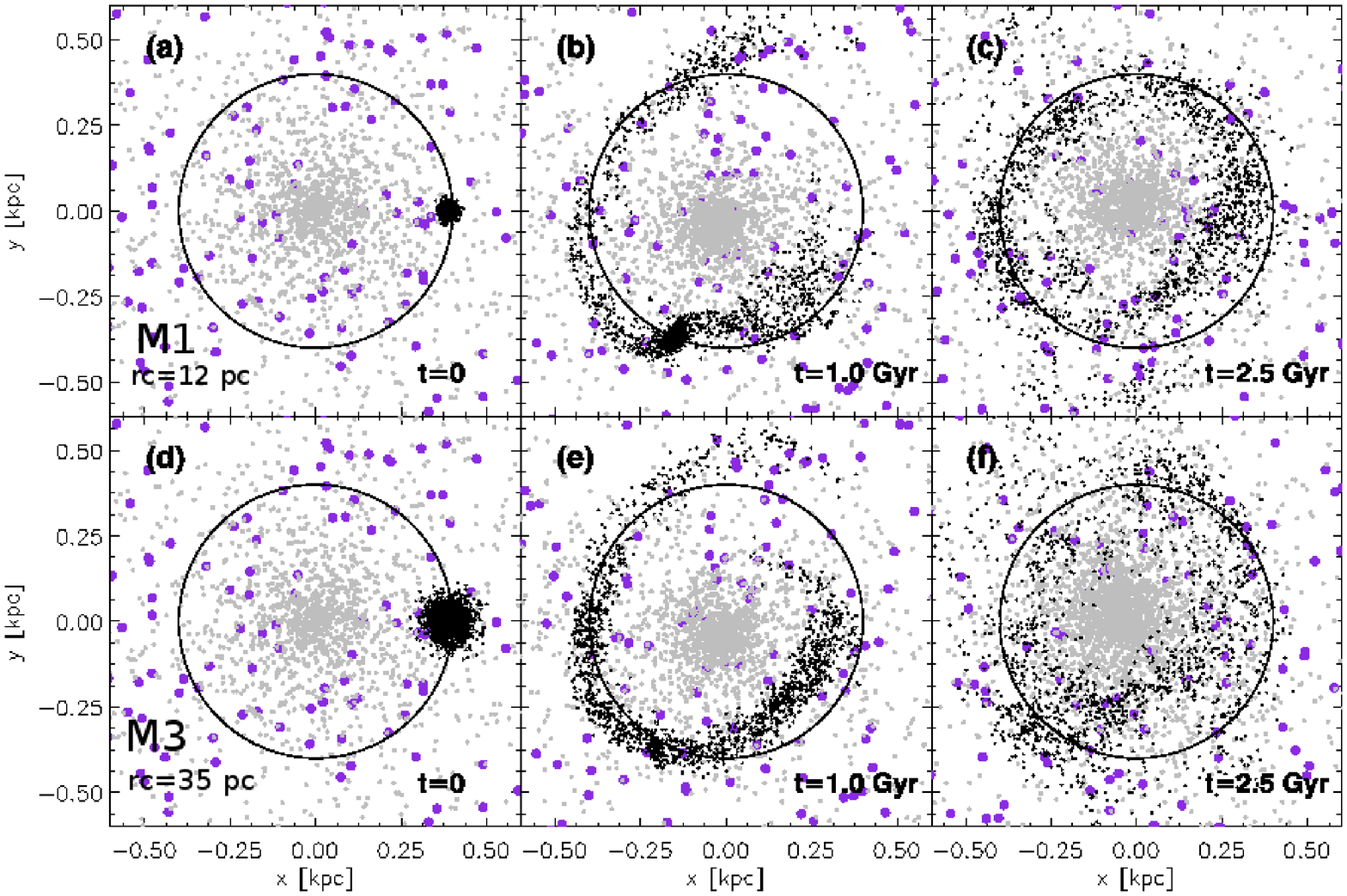}
\caption{Evolution of a clump embedded
in a smooth halo with a small core. The time is given at the bottom right
corner. 
The top panels ($a$, $b$ and $c$) show the case where the radius of the 
clump is $r_c=12$~pc (model $M1$). 
The bottom panels ($d$, $e$ and $f$) show the case where the radius of 
the clump is $r_c=35$~pc (model $M3$). 
The purple circles represent the soft halo DM particles; the small light 
gray circles indicate the extended stellar component;
and the black dots are for the particles of the stellar clump.}
\label{fig:res1}
\end{figure}

In Figure \ref{fig:res1} we show the evolution of the clump in the models $M1$ 
(clump radius $r_c=12$~pc) and $M3$ (clump radius $r_c=35$~pc)
for a halo with a small core ($r_{s}=0.91$~kpc).
At $t=1$~Gyr in model $M1$, the clump inflated its radius by a factor
$\sim1.5$ due to gravitational encounters with the halo particles, 
but it preserves its identity (see panel $(b)$ of
Figure \ref{fig:res1}). 
In model $M3$, the clump appears to be very disrupted 
already at $t=1$~Gyr (see panel $(e)$ in Figure \ref{fig:res1}). 
We can say
that, regardless the clump radius, the clump is totally destroyed 
at $t=2.5$~Gyr in both models (see panels $(c)$ 
and $(f)$ in Figure \ref{fig:res1}).

In order to quantify the destruction time of UMi's stellar clump in our
simulations, we built a map of the surface density of the stellar clump 
in the $(x,y)$-plane
at any given time $t$ in the simulation. We sample this two-dimensional map 
searching for the $10\times10$~pc size parcel that contains the highest mass (number of clump particles).
This region will be centered at the remnant of the clump. 
We define that a clump is destroyed when this region has reached a 
density of $1M_{\odot}$~pc$^{-2}$.
%\textbf{
When such a small value of the surface mass density is reached 
by UMi's clump, the column density of the clump is so low that
it would be unrecognisable from the underlying stellar component, 
and would be thus undetectable.
%}

\begin{figure}
\epsscale{.80}
\plotone{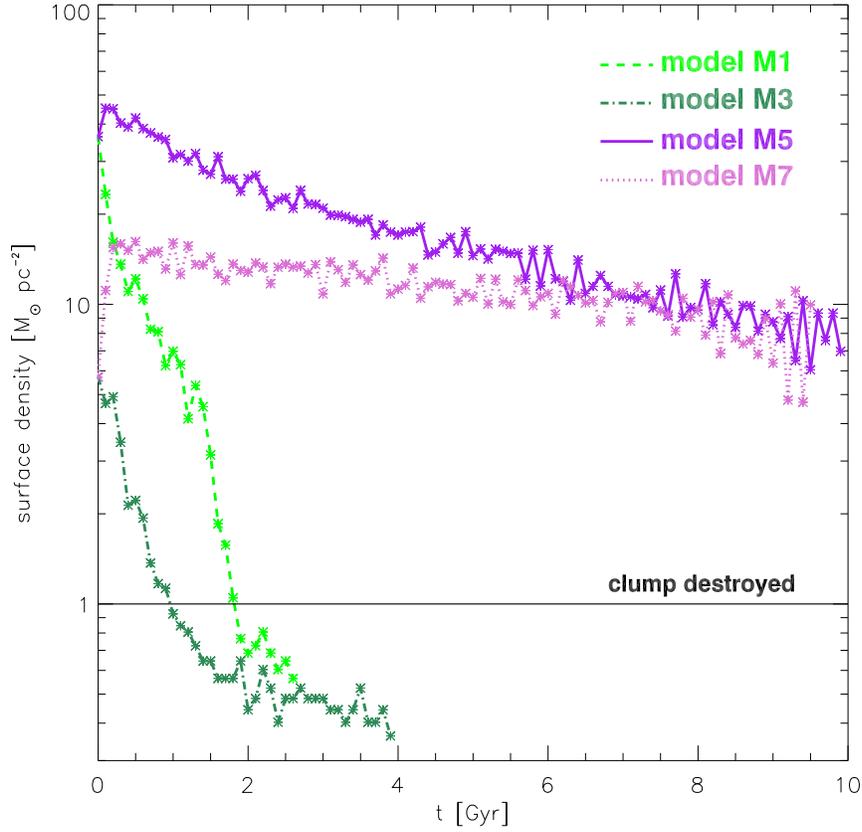}
 \caption{Surface density of the mass of UMi's clump map in the $(x,y)$-plane
at any given time $t$ in the simulation for the four models without substructure 
(see the models $M1$, $M3$, $M5$ and $M7$ in Table~\ref{tab:resultados1}). The
black line shows the destruction ($1 M_{\odot}$~pc$^{-2}$) line.}
 \label{fig:F3}
 \end{figure}

Figure \ref{fig:F3} shows the surface density of mass as a function of time, 
where the black line represents the surface density at which the clump 
is disrupted. 
The destruction times for both clumps in 
the small-core case are $1.8$ and $0.9$~Gyr for $r_c=12$~pc and $r_c=35$~pc, 
respectively (see last column of Table~\ref{tab:resultados1}). 
This implies that 
a halo core of $\sim0.4$~kpc is not large enough to guarantee 
the survival of the clump. This result is in agreement with \citeauthor{kleyna03}'s (\citeyear{kleyna03}) 
statement that if the core radius of the DM halo is equal to or smaller than the clump's 
orbit, the clump will get destroyed within $\sim 2$~Gyr. 
We recall that, in this case,
the DM halo has a core radius of $\sim0.4$~kpc and the clump has a 
galactocentric distance of $0.39$~kpc. 

\begin{figure*}
\epsscale{.80}
\plotone{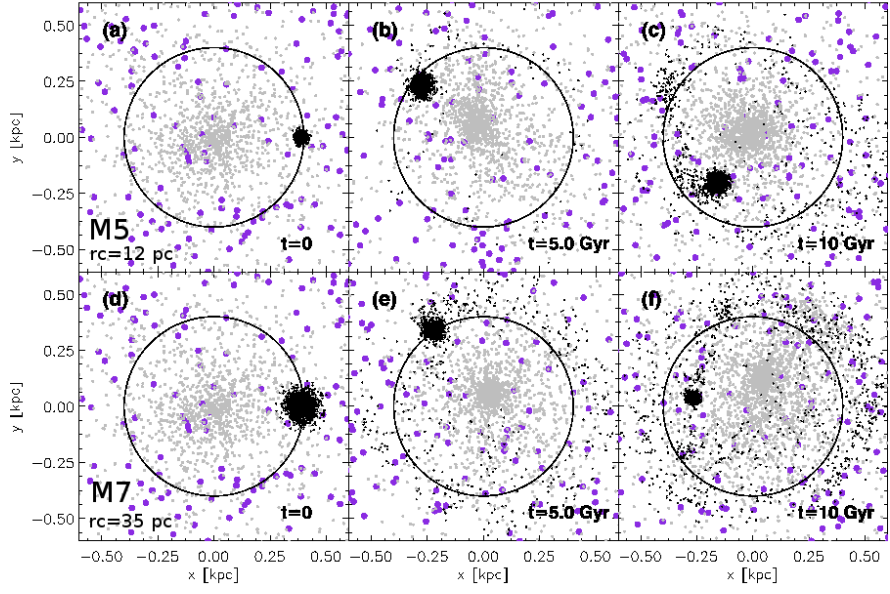}
\caption{ Same as Figure \ref{fig:res1} but for the big-core DM halo 
and integration times $t=0, 5$ and $10$~Gyr. The top panels show model
$M5$ and the bottom panels show model $M7$.
}
\label{fig:res2}
\end{figure*}

In Figure \ref{fig:res2} we show the evolution of the clump for both $r_{c}=12$~pc ($M5$) and 
$r_{c}=35$~pc ($M7$), embedded in a halo with a large core ($r_{s}=2.2$~kpc). 
In both cases
the clump survives for $\sim$ Hubble time. The small ($r_{c}=12$~pc) clump
expands up to $\sim 2$ times of its original size in the first $5$~Gyr 
(see panel $(b)$ of
Figure \ref{fig:res2}). After that, it maintains this size 
over one Hubble time. In the $M7$ model
($r_{c}=35$~pc), the clump loses some particles (see panel $(e)$ of
Figure \ref{fig:res2}) and slightly reduces its initial size (see panel $(e)$ of
Figure \ref{fig:res2}). Then the clump continues losing some particles and,
 as a consequence, it shrinks its initial size by a factor
of $\sim2.5$  by the end of the simulation ($\sim$ Hubble time), 
but remains undestroyed. The survival of the clump can also be seen in Figure 
\ref{fig:F3}; the surface mass densities in the $M5$ and $M7$ models 
lie well above the clump's destruction line (indicated by the horizontal black line).

This set of simulations tells us that 
a halo with a large core allows the survival of both clumps 
for $\sim$ Hubble time. This result reaffirms the belief that 
the UMi dSph galaxy should have a large core DM halo \citep{kleyna03},
instead of a cuspy DM density profile.

%%%%%%%%%%%%%%%%%%%%%%%%%%%%%%%%%%%%%%%%%%%%%%%%%

\subsection{Halo with substructure}
\label{sec:con_sub}
In this section we present the same simulations as those in 
\S\ref{sec:sin_sub} but
including substructure in the DM halo with $f=0.1$, 
$M_{max}=10^7M_{\odot}$ and
$M_{min}=10^{3.3}M_{\odot}$ (which is the maximum mass resolution for 
the DM halo particles). 
For the halo with a small core (where the total
mass of UMi is $M=2\times10^{9}M_{\odot}$), the number 
of substructure particles is of
$N_{sub}=5836$. For the halo with a large core 
($M=3\times10^{10}M_{\odot}$), we require $N_{sub}=22483$.

\begin{figure*}
\epsscale{.80}
\plotone{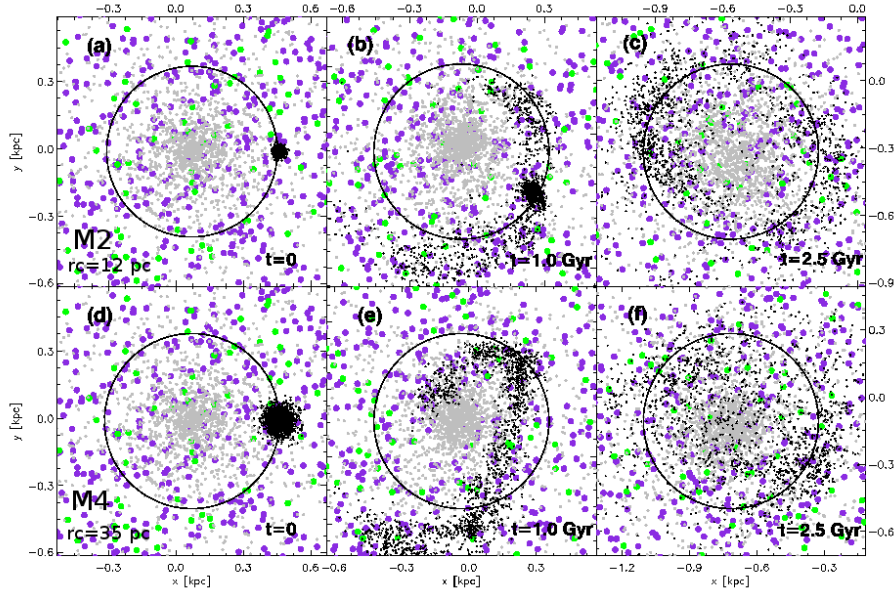}
\caption{Snapshots at  
$t=0, 1$ and $2.5$~Gyr for the small-core DM halo
with $10$\% of the DM halo's total mass in substructure (green circles).
The top panels show model $M2$ and the bottom panels show model $M4$.
The color code of the particles is the same as in Figure \ref{fig:res1}.}
\label{fig:M2M4}
\end{figure*}

\begin{figure}
\epsscale{.80}
\plotone{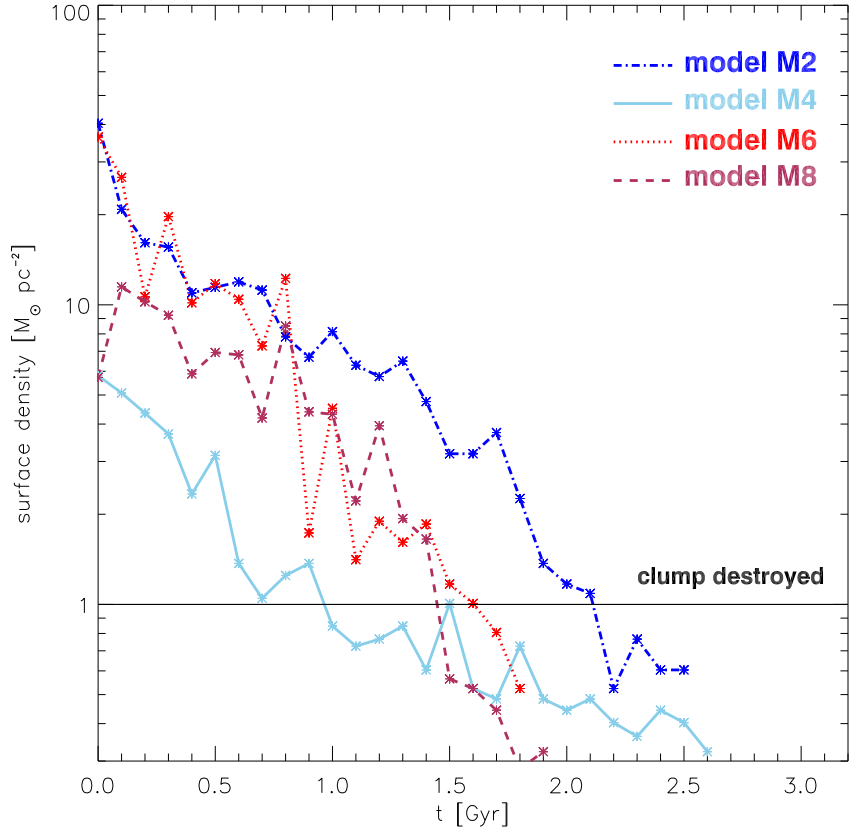}
 \caption{Same as Figure \ref{fig:F3} for the four models
with dark-matter subhalos
(see models $M2$, $M4$, $M6$ and $M8$ in Table~\ref{tab:resultados1})
}
 \label{fig:F4}
 \end{figure}

In Figure~\ref{fig:M2M4} we show snapshots of the models $M2$ and $M4$ 
(small-core clumpy DM halo)
at $t=0,1$ and $2.5$~Gyr. The purple circles represent the DM particles, 
the small light gray circles represent UMi's  extended stellar component particles, 
the black points represent the particles that 
make up the stellar clump and in green we show the substructure particles. 
From Figure \ref{fig:M2M4} (see panels $a$, $b$ and $c$), we can see that 
the small ($r_c=12$~pc) 
clump has increased its size by a factor of $\sim2$ after $1$ Gyr, 
slightly larger compared to the
case without substructure. 
%\textbf{
We built a map of the surface density of the stars initially in
the clump (see Figure \ref{fig:F4}), 
similar as that in the case without substructure.
%}
%where the clump appears to be
A clump with an initial radius of $r_c=35$~pc survives $\sim1$~Gyr,
which is also approximately the same time as in the non-substructured case 
(see last column of Table \ref{tab:resultados1}).
These results imply that the dissolution by tidal forces in a DM halo with a
small core plays a major 
role in the dynamical evolution of the clump, more than the 
interaction between the massive substructure particles and the 
particles of the clump.

\begin{figure}
\epsscale{.80}
\plotone{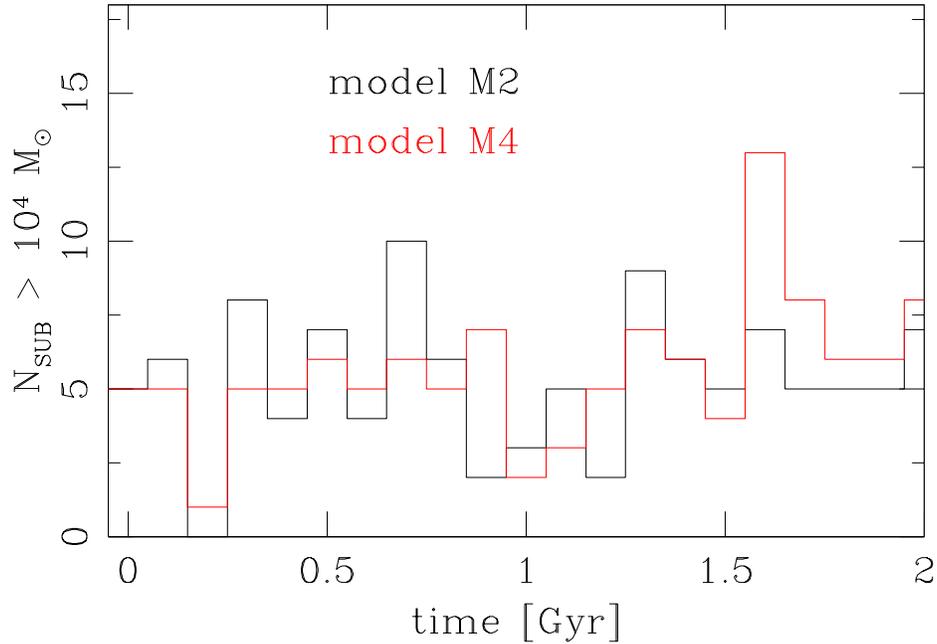}
%\plotone{hist_M2_M4.eps}
\caption{Number of halo particles with mass $>10^{4}M_{\odot}$,
inside a $0.39$~kpc sphere around the center of density of the simulation (i.e. inside the clump's orbit) for model $M2$ (black line) 
for model $M4$ (red line). 
The number of particles was computed at each $0.1$~Gyr.}
\label{fig:espectro}
\end{figure}

In Figure~\ref{fig:espectro} we show the number of  
substructure particles with masses $>10^{4}M_{\odot}$ inside 
a sphere of $0.39$-kpc radius   
for models $M2$ and $M4$.
This radius corresponds to the radius at which the clump is orbiting, 
and thus massive subhalos within that radius will 
have a major effect in the clump's destruction. 
In these models, the number of particles more massive 
than $10^4M_{\odot}$, inside the clump's orbit, ranges from $1$ to $9$,
with a mean number of $5$ particles.

\begin{figure*}
\epsscale{.80}
\plotone{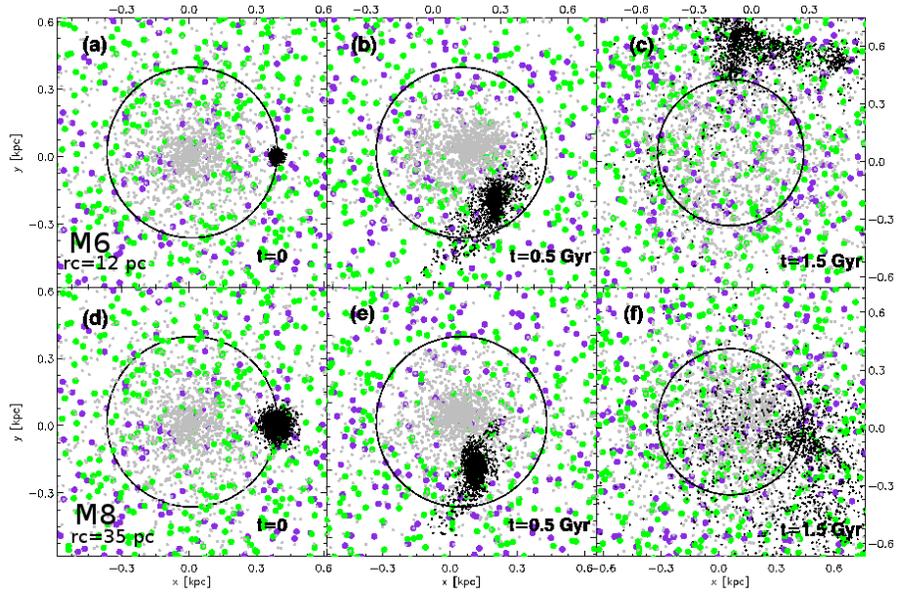}
\caption{Same as Figure \ref{fig:M2M4} but at times $t=0, 0.5$ and $1.5$~Gyr.
The top panels show model $M6$ and the bottom panels show model $M8$.
}
\label{fig:M6M8}
\end{figure*}

\begin{figure}
\epsscale{.80}
\plotone{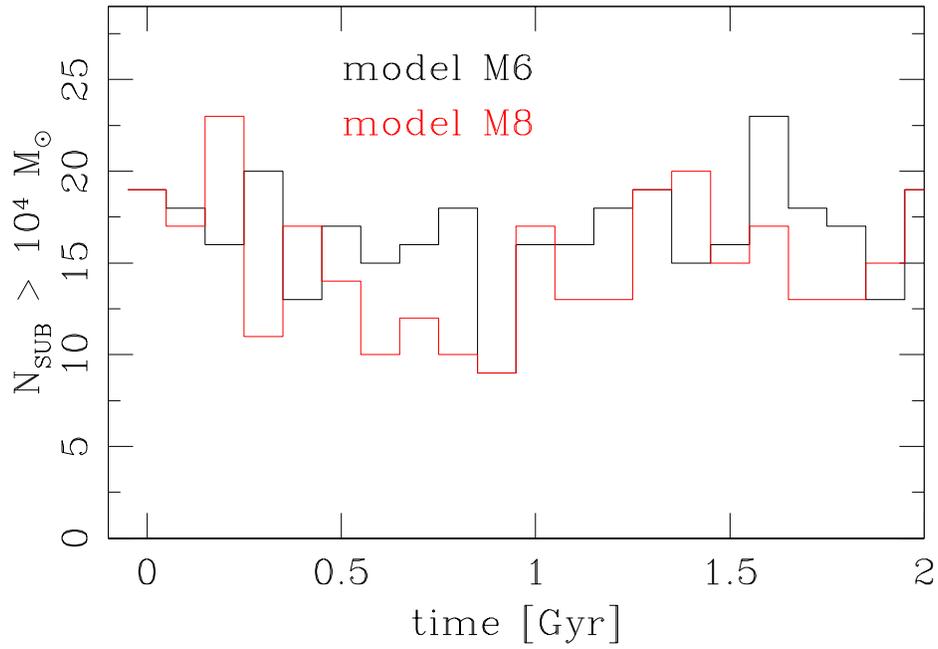}
\caption{ Same as Figure \ref{fig:espectro} but for models $M6$ (black) and
$M8$ (red).}
\label{fig:espectro_big}
\end{figure}

In Figure~\ref{fig:M6M8} we show snapshots for the clumpy halo with a big core
(models $M6$ and $M8$) at $t=0,0.5$ and $1.5$~Gyr. In 
both simulations, the clump appears to be enlarged 
at $t=0.5$~Gyr (see panels $b$ and $e$ of 
Figure~\ref{fig:M6M8}) and, at $t=1.5$~Gyr, 
it is practically disrupted. 
The clump is dissolved at $1.6$~Gyr in model $M6$, and 
at $1.4$ Gyr in model $M8$.  The number of
substructure particles with mass larger than $10^4M_{\odot}$ is shown
in Figure~\ref{fig:espectro_big}. In the case of a halo with a big core,
the number of particles with masses $>10^4M_{\odot}$ 
inside a $0.39$~kpc sphere ranges between
$8$ and $23$,  with a mean number of $16$. 
In the models with a large core, but without substructure, the 
clumps ($12$ and $35$~pc) remain undestroyed for $\sim$ Hubble time, which
means that the destruction of the clump in the clumpy halo case 
is due to the
random walk in momentum space that the stars in the clump undergo
by the collisions with the substructure particles. This effect is so 
important that even with a core as large as $\sim1$~kpc,
the clump does not manage to survive the continuous encounters with the massive
substructure particles of the DM halo.

%%%%%%%%%%%%%%%%%%%%%%%%%%%%%%%%%%%%%%%%%%%%%%%%%%%%%%%%%%%%%%%%%%%%%%%%%%%%%%%%%%%%%%%%%%%%%%%%

\subsection{A comparison with analytical estimates}
\citet{sanchez07} studied the abundance of Very Massive Objects (VMO) 
in the DM halo of Fornax and UMi. 
These VMOs can be compared with the subhalos that
we have studied in this paper for UMi. 
In the impulse approximation, they found that, 
if the progenitor cluster  became unbound immediately after formation,
the mass of the VMOs should be
\begin{equation}
 M_h \leq 2.5\times10^3 M_{\odot} \mbox{ } 
\left(\frac{f\rho_{\rm dm}}{0.1 M_{\odot}\mbox{ pc}^{-3}}\right)^{-1} 
\left(\frac{\sigma_{\rm dm}}{20 \mbox{ km s}^{-1}} \right)
\end{equation}
in order to have a clump as dynamically cold as observed.
Here $\rho_{\rm dm}$ and $\sigma_{\rm dm}$ are the density and velocity
dispersion of the dark matter particles in the UMi halo. 
For $f=0.1$, this upper limit implies $M_h \leq 2.5\times 10^4 M_{\odot}$.
Therefore, the maximum number of VMO within the orbital radius $r_{\rm orb}$
is $\simeq 4\pi f r_{\rm orb}^{3}\rho_{\rm dm}/(3M_{h})\simeq 100$.
In the case of a mass spectrum, the maximum number of subhalos with
masses $\leq 2.5\times 10^4 M_{\odot}$ are expected to be smaller because
the dissolution time of the clump due to collisions
with subhalos of mass $M$ goes as $\propto M^{2}dN(M)$.
Our $N$-body simulations have shown that, if the subhalos follow
a mass spectrum and if self-gravity of the clump is taken into account,
$\sim 16$ subhalos
with masses larger than $10^{4}M_{\odot}$ are enough to disrupt the
clump if the tidal perturbation by the dark halo is also included.

%%%%%%%%%%%%%%%%%%%%%%% Section conclusions %%%%%%%%%%%%%%%%%%%%%%%%%%%%%%%%%%%%%%%%%%%%%

\section{Concluding remarks} 
\label{sec:conclusions}
We studied the dynamical consequences of the putative substructure 
in UMi's DM halo in the stellar component.
We ran $N$-body simulations of a stellar clump orbiting
in a live cored DM halo.
If the distribution of mass in the dark halo is smooth, 
a dark halo with a scale length of $0.91$ kpc, which corresponds
to a core of $\sim 0.4$ kpc, cannot preserve the integrity
of the clump.
On the other hand, for a dark halo with a big core (scale length of $2.2$~kpc,
core radius of $\sim 1$ kpc),
the clump survives
for approximately a Hubble time \citep{kleyna03}.

When $10\%$ of the original DM halo total mass resides in
compact subhalos, 
the clump dissolves in roughly the same timescale in the small core case.
This means that the dissolution effects by tidal forces by the
small-core DM halo are greater than due to the gravitational
scattering with the massive 
substructure particles.

In the big-core case there is a large number of particles in substructure 
($N_{sub}=22483$)
with a mean number of $16$ particles inside a sphere of 
radius $0.39$~kpc (the radius of the orbit of the clump) with
masses greater than $10^4 M_{\odot}$ throughout the simulation.
The effect of the substructures over the clump in this case 
results in the complete destruction of the clump within $\sim1.6$~Gyr
for the clump with $r_c=12$~pc and within $\sim1.4$~Gyr
for the clump with $r_c=35$~pc.
In a smooth dark-matter halo,
the large DM core halo ensures the longevity of the clump for almost
one Hubble time, but in a clumpy halo, the clump
is erased because of the
random walk in momentum space that the stars in the clump undergo
by collisions with the massive substructure particles.

It remains to study a more realistic scenario where the subhalos are not point
particles (in which case resembles more the case of VMOs) but are 
extended perturbers. It would be worthwhile to quantify the effect that
these extended subhalos would imprint compared to the point mass case 
studied here, and see in a more general case if subhalos in dSph galaxies 
impose a strong test to the $\Lambda$CDM scenario.
Preliminary simulations suggest that the disruption timescale of
the clump increases by $60\%$--$70\%$ if subhalos are extended, which is not
enough to account for the survival of the clump. A thorough
study will be presented elsewhere.

%%%%%%%%%%%%%%%%%%%%%%%%%%%%%%%%%%%%%%%%%%%%%%%%%%%%%%%%%%%%%%%%%%%%%%%%%%%%%%%%%%%%%%%%%

\acknowledgments

V.L. gratefully acknowledges support from the Alexander von Humboldt 
Foundation fellowship, and the anonymous referee for very useful comments 
that improved the presentation of the paper.. 
A.J. and E.K.G. acknowledge support from the Collaborative Research Center ``The Milky Way System"
(SFB881) of the German Research Foundation (DFG), 
especially subprojects A1 and A2. F.J.S.S.~was partly supported by
CONACyT project 165584 and PAPIIT project IN106212. 

%and an anonymous referee for very useful comments that improved
%the presentation of the paper.

%%%%%%%%%%%%%%%%%%%%%%%%%%%%%%%%%%%%%%%%%%%%%%%%%%%%%%%%%%%%%%%%%%%%%%%%%%%%%%%%%%%%%%%%%%%%

%%%%%%%%%%%%%%%%%%%%%%%%%%%%%%%%%%%%%%%%%%%%%%%%%%%%%%%%%%%%%%%%%%%%%%%%%%%%%%

\clearpage



\clearpage

\begin{table}
\begin{center}
\caption{Parameters used in the simulations. Here $r_{s}$ accounts
for the DM halo scale length; M is the total mass of the halo, and $r_{c}$ is the radius 
of the stellar clump in UMi.}
 \begin{tabular}{@{}ccccccccc@{}}
\tableline\tableline
Model & sub- & $r_{s}$ & M & $r_{c}$ & destruction time &\\
 & structure & [kpc]  & [$M_{\odot}$] & [pc] & [Gyr]  & \\
\tableline
M1 & no & $0.91$   &$2\times10^9$ & $12$ & $\sim 1.8$& \\

M2& yes & $0.91$   &$2\times10^9$ & $12$ &  $\sim 2.1$& \\

M3 & no & $0.91$   &$2\times10^9$ & $35$ & $\sim 0.9$& \\

M4& yes & $0.91$   &$2\times10^9$ & $35$ & $\sim 1$& \\

M5& no & $2.2$   &$3\times10^{10}$ & $12$ & $>10$& \\

M6& yes & $2.2$   &$3\times10^{10}$ & $12$ & $\sim 1.6$& \\

M7& no & $2.2$   &$3\times10^{10}$ & $35$ & $>10$& \\

M8& yes & $2.2$   &$3\times10^{10}$ & $35$ & $\sim 1.4$& \\
\tableline
\end{tabular}
\label{tab:resultados1}
\end{center}
\end{table}

\clearpage

\begin{table}
%\begin{sidewaystable}
\begin{center}
\caption{Setup properties of the particles in the simulations. Total mass, 
mass per particle and number of particles of the three UMi modeled components, the
radii of the grids and the resolution $l$ of the middle grid (for $N_c=128$) 
are shown.}
\begin{tabular}{@{}ccccccccc@{}}
\tableline\tableline
Component & M & m$_{p}$& N & inner grid & middle grid & $l_{128}$&\\% $l_{256}$&\\
 &[$M_{\odot}$] &[$M_{\odot}$]& $10^6$ & [kpc]  & [kpc] & [kpc] &\\% [kpc]&\\
\tableline 
Halo with small core& $2\times10^9$ & $2\times10^3$& $1$ & $1$ & $20$ & $0.322$&\\%$0.158$&\\

Halo with large core& $3\times10^{10}$ & $2\times10^3$&$15$&$2$& $30$ & $0.483$&\\%$0.237$&\\

Extended stellar component& $9\times10^5$ & $9$ & $0.1$&$1$ & $10$ & $0.161$&\\%$0.079$&\\

Clump with $r_{c}=12$ pc & $4.03\times10^4$ & $4.03$ &$0.01$& $0.012$& $0.12$& $0.0019$&\\% $9.48\times10^4$& \\

Clump with $r_{c}=35$ pc & $4.03\times10^4$ & $4.03$ &$0.01$& $0.035$& $0.35$& $0.0056$&\\%$0.00276$&\\

\tableline
\end{tabular}
\label{tab:code}
\end{center}
\end{table}
%\end{sidewaystable}


\begin{thebibliography}{99}

\bibitem [\protect\citeauthoryear{Agnello \& Evans}{2012}]{agn12}
Agnello, A., \& Evans, N. W. 2012, arXiv:1205.6673

\bibitem [\protect\citeauthoryear{Amorisco et al.}{2011}]{amorisco}
Amorisco, N. C. \& Evans, N. W. 2012, MNRAS, 419, 184

\bibitem [\protect\citeauthoryear{Avila-Reese et al.}{2001}]{avi01}
Avila-Reese, V., Col\'{\i}n, P., Valenzuela, O., D'Onghia, E., \&  Firmani, C.
2001, ApJ, 559, 516

\bibitem[\protect\citeauthoryear{Battaglia et al.}{2008}]{battaglia}
Battaglia, G., Helmi, A., Tolstoy, E., Irwin, M., Hill, V. \& Jablonka, P., 
2008, ApJ, 681, L13 

\bibitem[\protect\citeauthoryear{Benson et al.}{2002}]{ben02}
Benson, A. J., Frenk, C. S., Lacey, C. G., Baugh C. M., \& Cole S. 2002,
MNRAS, 333, 177

\bibitem[\protect\citeauthoryear{Binney \& Tremaine}{2008}]{binney}
Binney J. \& Tremaine S. 2008, Galactic Dynamics, 2nd edn. Princeton Univ.
Press, Princeton, NJ

\bibitem[\protect\citeauthoryear{Bovill \& Ricotti}{2011}]{bovill}
Bovill, M. S. \& Ricotti, M., 2011, ApJ, 741, 18

\bibitem[\protect\citeauthoryear{Boylan-Kolchin et al.}{2011}]{boylan}
Boylan-Kolchin, M., Bullock, J. S. \& Kaplinghat, M., 2011, MNRAS, 415, 40

\bibitem[\protect\citeauthoryear{Carlberg}{2009}] {carlberg09}
Carlberg, R. G. 2009, ApJ, 705, L223

\bibitem[\protect\citeauthoryear{Carlberg et al.}{2011}] {carlberg11}
Carlberg R. G., et al., 2011, ApJ, 731, 124

\bibitem[\protect\citeauthoryear{Carrera et al.}{2002}] {carrera02}
Carrera, R., Aparicio, A., Mart\'{\i}nez-Delgado, D., Alonso-Garc\'{\i}a, J.,
2002, AJ, 123, 3199

\bibitem[\protect\citeauthoryear{Cen}{2000}]{cen00}
Cen, R. 2001, ApJ, 546, L77

\bibitem[\protect\citeauthoryear{Chen et al.}{2010}]{chen10}
Chen, D. \& McGaugh, S. 2010, RAA, 10, Issue 12, 1215

%\bibitem[\protect\citeauthoryear{C\^{o}t\'{e} et al.}{1999}]{cote99}
%C\^{o}t\'{e}, P., Mateo, M., Olszewski, E. \& Cook, K. 1999, ApJ, 526, 147

\bibitem[\protect\citeauthoryear{de Blok et al.}{2001}]{deblok01}
de Blok, W. J. G., McGaugh, S. S., Bosma, A., \& Rubin, V. C. 2001, ApJ, 552, L23

\bibitem[\protect\citeauthoryear{Dehnen}{1993}]{dehnen93}
Dehnen, W. 1993, MNRAS, 265, 250

\bibitem[\protect\citeauthoryear{Diemand et al.}{2008}]{diemand08}
Diemand, J., Kuhlen, M., Madau, P., Zemp, M., Moore,
B., Potter, D., \& Stadel, J. 2008, Nature, 454, 735

\bibitem[\protect\citeauthoryear{El-Zant, Shlosman \& Hoffman}{2001}]{elz01}
El-Zant, A., Shlosman, I., \& Hoffman, Y. 2001, ApJ, 560, 636

\bibitem[\protect\citeauthoryear{Fellhauer et al.}{2000}]{fellhauer00}
Fellhauer, M., Kroupa, P., Baumgardt, H., Bien, R., Boily, C. M., Spurzem, R. \& Wassmer, N.,
2000, New Astron., 5, 305

\bibitem[\protect\citeauthoryear{Font et al.}{2011}]{fon11}
Font, A. S. et al. 2011, MNRAS, 417, 1260

\bibitem[\protect\citeauthoryear{Gao et al.}{2004}]{gao04}
Gao L., White S. D. M., Jenkins A., Stoehr F., Springel V., 2004, MNRAS, 355, 819

\bibitem[\protect\citeauthoryear{Gilmore et al.}{2007}]{gilmore07}
Gilmore, G., Wilkinson, M. I., Wyse, R. F. G., Kleyna, J. T., Koch, A. et al. 2007, AJ, 663, 94

\bibitem[\protect\citeauthoryear{Goerdt et al.}{2006}]{goerdtd06}
Goerdt, T., Moore, B., Read, J. I., Stadel, J. \& Zemp, M., 2006, MNRAS, 368, 1073

\bibitem[\protect\citeauthoryear{Goerdt et al.}{2010}]{goerdtd10}
Goerdt, T., Moore, B., Read, J. I., Stadel, J. \& Zemp, M., 2010, ApJ, 725, 1707

\bibitem[\protect\citeauthoryear{Governato et al.}{2010}]{gov10}
Governato, F. et al. 2010, Nature, 463, 203

\bibitem[\protect\citeauthoryear{Governato et al.}{2012}]{gov12}
Governato, F. et al. 2012, MNRAS, 422, 1231 

\bibitem[\protect\citeauthoryear{Grebel et al.}{2003}]{grebel03}
Grebel, E. K., Gallagher, J. S., \& Harbeck, D., 2003, AJ, 125,1924

\bibitem[\protect\citeauthoryear{Green et al.}{2005}]{green05}
Green, A. M.,  Hofmann, S., \& Schwarz, D. J., 2005, JCAP,
08, 003

\bibitem[\protect\citeauthoryear{Hofmann et al.}{2001}]{hofmann01}
Hofmann, S., Schwarz, D. J., \& St\"ocker, H. 2001, Phys.~Rev.~D, 64, 083507

\bibitem[\protect\citeauthoryear{Jardel \& Gebhardt}{2012}]{jardel12}
Jardel, J. R., \& Gebhardt, K. 2012, ApJ, 746, 89

\bibitem[\protect\citeauthoryear{Jin et al.}{2005}]{jin05}
Jin, S., Ostriker, J. P., \& Wilkinson, M. I. 2005, MNRAS, 359, 104

\bibitem[\protect\citeauthoryear{Johnston et al.}{2002}]{johnston02}
Johnston, K. V., Spergel, D. N., \& Haydn, C. 2002, ApJ, 570, 656

\bibitem[\protect\citeauthoryear{Kannan et al.}{2012}]{kannan12}
Kannan, R., Maccio, A. V., Pasquali, A., Moster, B. P. \& Walter, F. 2012, ApJ, 746, 10

\bibitem[\protect\citeauthoryear{Kaplinghat, Knox \& Turner}{2000}]{kap00}
Kaplinghat, M., Knox, L., \& Turner, M. S. 2000, Physical Review Lett.,
85, 3335

\bibitem[\protect\citeauthoryear{Khoperskov et al.}{2007}]{khoperskov}
Khoperskov, A. V., Just, A., Korchagin, V. I., Jalali, M. A., 2007,
A\&A, 473, 31 

\bibitem[\protect\citeauthoryear{Kleyna et al.}{1998}]{kleyna98}
Kleyna, J. T., Geller, M. J., Kenyon, S. J., Kutz, M. J., Thorstensen, J. R., 1998, ApJ, 115, 2359

\bibitem[\protect\citeauthoryear{Kleyna et al.}{2003}]{kleyna03}
Kleyna, J. T., Wilkinson, M. I., Gilmore, G. \& Evans, N. W., 2003, ApJ, 588, L21

\bibitem[\protect\citeauthoryear{Klypin et al.}{1999}]{klypin}
Klypin, A., Kravtsov, A., Valenzuela, O. \& Prada, F., 1999, ApJ, 522, 82

\bibitem[\protect\citeauthoryear{Kniazev et al.}{2009}]{kniazev}
Kniazev, A. Y., Brosch, N., Hoffman, G. L., Grebel, E. K., Zucker, D. B. \& Pustilnik, S. A.
2009, MNRAS, 400, 2054

\bibitem[\protect\citeauthoryear{Lora et al.}{2009}]{lora}
Lora, V., S\'anchez-Salcedo, F. J., Raga, A. C. \& Esquivel, A., 2009, ApJ, 699, L113

%\bibitem[\protect\citeauthoryear{Mateo et al.}{1993}]{mateo93}
%Mateo, M., Olszewski, E. W., Pryor, C., Welch, D. L. \& Fischer, P. 1993, AJ, 105, 510

\bibitem[\protect\citeauthoryear{Ma et al.}{2004}]{ma04}
Ma, C.-P., \& Boylan-Kolchin, M. 2004, Phys. Rev. Lett., 93, 021301

\bibitem[\protect\citeauthoryear{Macci\`o et al.}{2012}]{mac12}
Macci\`o, A. V., Stinson, G., Brook, C. B., Wadsley, J., Couchman, H. M. P., 
Shen, S., Gibson, B. K., \& Quinn, T. 2012, ApJ, 744, L9

\bibitem[\protect\citeauthoryear{Mashchenko, Couchman \& Wadsley}{2006}]{mas06}
Mashchenko, S., Couchman, H. M. P., \& Wadsley, J. 2006, Nature, 442, 539 

\bibitem[\protect\citeauthoryear{Mateo}{1998}]{mateo98}
Mateo, M. L., 1998, ARA\&A, 36, 435

\bibitem[\protect\citeauthoryear{Mo \& Mao}{2004}]{mo04}
Mo, H. J., \& Mao, S. 2004, MNRAS, 353, 829

\bibitem[\protect\citeauthoryear{Moore et al.}{1999}]{moore99}
Moore, B., Ghigna, S., Governato, F., Lake, G., Quinn, T., Stadel, J., \& Tozzi,
P. 1999, ApJ, 524, L19

\bibitem[\protect\citeauthoryear{Navarro et al.}{1996}]{nfw}
Navarro J. F., Frenk C. S. \& White S. D. M., 1996, ApJ, 462, 563

\bibitem[\protect\citeauthoryear{Odenkirchen et al.}{2001}]{odenkirchen}
Odenkirchen, M., Grebel, E. K., Harbeck, D., Dehnen, W., Rix, H. W., et al.
2001, AJ, 122, 2538

\bibitem[\protect\citeauthoryear{Ostriker et al.}{2003}]{ostriker}
Ostriker, J. P. \& Steinhardt, P., 2003, Science, 300, 1909

\bibitem[\protect\citeauthoryear{Palma et al.}{2003}]{palma03}
Palma, C., Majewski, S. R., Siegel, M. H., Patterson, R. J., Ostheimer, J. C. \& Link, R., 
2003, AJ, 125, 1352

\bibitem[\protect\citeauthoryear{Pe\~narrubia et al.}{2010}]{penarrubia10}
Pe\~narrubia, J., Koposov, S. E., Walker, M. G., Gilmore, G., Wyn E. N. \& Mackay, C. D.
2010, [arXiv:1005.5388]

\bibitem[\protect\citeauthoryear{Pe\~narrubia et al.}{2008}]{penarrubia08}
Pe\~narrubia, J., Navarro, J. F. \& McConnachie, A. W., 
2008, ApJ, 673, 266

\bibitem[\protect\citeauthoryear{Pontzen \& Governato}{2012}]{pon12}
Pontzen, A., \& Governato, F. 2012, MNRAS, 421, 3464

\bibitem[\protect\citeauthoryear{Read et al.}{2006}] {read06}
Read, J. I., Wilkinson, M. I., Evans, N. W., Gilmore, G., Kleyna, J. T., 2006, MNRAS, 367, 387 

\bibitem[\protect\citeauthoryear{Rocha et al.}{2011}] {rocha11}
Rocha, M., Peter, A. H. G., Bullock, J. S., 2011, arXiv:1110.0464

\bibitem[\protect\citeauthoryear{Romano-D\'iaz et al.}{2008}] {romano08}
Romano-D\'iaz, E., Shlosman, I., Heller, C. \& Hoffman, Y., 2008, ApJ, 687, L13

\bibitem[\protect\citeauthoryear{Romano-D\'iaz et al.}{2009}] {romano09}
Romano-D\'iaz, E., Shlosman, I., Heller, C. \& Hoffman, Y., 2009, ApJ, 702,
1250

\bibitem[\protect\citeauthoryear{S\'anchez-Salcedo}{2003}]{san03}
S\'anchez-Salcedo, F. J. 2003, ApJ, 591, L107 

\bibitem[\protect\citeauthoryear{S\'anchez-Salcedo et al.}{2006}]{sanchez1}
S\'anchez-Salcedo, F. J., Reyes-Iturbide, J. \& Hernandez, X., 2006, MNRAS, 370, 1829.

\bibitem[\protect\citeauthoryear{S\'anchez-Salcedo \& Lora }{2007}]{sanchez07}
S\'anchez-Salcedo, F. J., \& Lora, V., 2007, ApJ, 658, L83 

\bibitem[\protect\citeauthoryear{S\'anchez-Salcedo \& Lora }{2010}]{sanchez10}
S\'anchez-Salcedo, F. J., \& Lora, V., 2010, MNRAS, 407, 1135 

\bibitem[\protect\citeauthoryear{Somerville}{2002}]{som02}
Somerville, R. S. 2002, ApJ, 572, 23

\bibitem[\protect\citeauthoryear{Shu et al.}{1987}]{shu}
Shu, F. H., Adams, F. C., Lizano, S., 1987, ARA\&A, 25, 23

\bibitem[\protect\citeauthoryear{Spergel \& Steinhardt}{2000}]{spe00}
Spergel, D. N., \& Steinhardt, P. J. 2000, Physical Review Lett., 84,
3760

\bibitem[\protect\citeauthoryear{Springel et al.}{2008}]{springel08}
Springel, V., Wang, J., Vogelsberger, M., Ludlow, A., Jenkins, A., Helmi, A., 
Navarro, J. F., Frenk, C. S. \& White, S. D. M., 2008, MNRAS, 391 1685

\bibitem[\protect\citeauthoryear{Strigari et al.}{2007}] {strigari07}
Strigari, L. E., Bullock, J. S., Kaplinghat, M., Diemand, J., Kuhlen, K \& Madau, P.,
2007, ApJ, 669, 676

\bibitem[\protect\citeauthoryear{Strigari et al.}{2008}] {strigari08}
Strigari, L. E., Bullock, J. S., Kaplinghat, M., Simon, J. D., Geha, M., Willman, B. \& Walker, M. G., 
2008, Nature, 454, 1096

\bibitem[\protect\citeauthoryear{Totani}{2010}] {totani10}
Totani, T. 2010, PASJ, 62, L1 

\bibitem[\protect\citeauthoryear{Tremaine et al.}{1994}] {tremaine94}
Tremaine, S., Richstone, D. O., Byun, Y.I., et al. 1994, AJ, 107, 634

\bibitem[\protect\citeauthoryear{Walker et al.}{2009}] {walker09}
Walker, M. G., Mateo, M., Olszewski, E. W., Pe\~narrubia, J., Wyn E. N. \& Gilmore, G.
2009, ApJ, 704, 1274

\bibitem[\protect\citeauthoryear{Walker \& Pe\~narrubia}{2011}] {walker11}
Walker, M. G., \& Pe\~narrubia, J. 2011, ApJ, 742, 20

\bibitem[\protect\citeauthoryear{Widrow}{2000}] {widrow00}
Widrow, L. M. 2000, ApJS, 131, 39

\bibitem[\protect\citeauthoryear{Wilkinson et al.}{2004}]{wilkinson04}
Wilkinson, M. I., Kleyna, J. T., Evans, N. W., Gilmore, G. F., Irwin, M. J. \& Grebel, E. K., 2004, ApJ, 611, L21

\bibitem[\protect\citeauthoryear{Yoon et al.}{2011}]{yoon11}
Yoon, J. H., Johnston, K. V., \& Hogg, D. W. 2011, ApJ, 731, 58 

\bibitem[\protect\citeauthoryear{Young}{1999}]{young99}
Young, L. M., 2000, AJ, 117, 1758

\bibitem[\protect\citeauthoryear{Young}{2000}]{young00}
Young, L. M., 2000, AJ, 119, 188

\end{thebibliography}
\end{document}